\documentclass{ws-procs961x669}            
\usepackage{epsfig}         
\newcommand \bel   {\begin{equation}\label}
\newcommand \bei  {\begin{itemize}}
\newcommand \eei  {\end{itemize}}

\newcommand \Tbb {\mathbb T}
\newcommand \gmoi {g^-}
\newcommand \Kmoi {K^-} 
\newcommand \phimoi {\phi^-} 
 
\newcommand \gpoi {g^+}
\newcommand \Kpoi {K^+} 
\newcommand \phipoi {\phi^+} 
 
\newcommand \Tr {\textbf{Tr}}
\newcommand \Ibf {\textbf{I}}
\newcommand \Sbf {\textbf{S}}
\newcommand \Div {\textbf{Div}} 
\newcommand \Mcal {\mathcal M} 
\newcommand \Kcirc {\mathring{K}}

\newcommand \Hcal {\mathcal H}
\newcommand \del   {{\partial}} 
\newcommand \eps   {\varepsilon}     
\newcommand \RR   	{{\mathbb R}}   
\newcommand \be {\begin{equation}}
\newcommand \ee {\end{equation}}

\begin{document}
\title{Gravitational singularities, scattering maps for bouncing,   
\\
and structure-preserving algorithms\footnote{Proceedings of the Sixteenth Marcel Grossmann Meeting (MG16), July 2021.}
}

\author{Philippe G. LeFloch}

\address{Laboratoire Jacques-Louis Lions and Centre National de la Recherche Scientifique
\\
Sorbonne Universit\'e, 4 Place Jussieu, 75252 Paris, France. 
\\
Email: {\sl contact@philippelefloch.org}.
Blog: philippelefloch.org
\\
\vskip.3cm
\bf October 2021
} 

\begin{abstract} This note emphasizes the role of multi-scale wave structures and junction conditions 
in many fields of physics,
 from the dynamics of fluids with non-convex equations of state 
to the study of gravitational singularities and bouncing cosmologies in general relativity. 
Concerning the definition and construction of bouncing spacetimes, we review the recent proposal in collaboration with B. Le Floch and G. Veneziano based on the notion of singularity scattering maps. We also present recent numerical investigations of small-scale phenomena arising in compressible fluid flows on FRLW or Kasner geometries for which we developed structure-preserving algorithms. 
\end{abstract}

 

\bodymatter 

\section{Introduction}
\label{sect-1}

Recent advances on the mathematical modeling of complex fluid flows containing nonlinear waves and gravitational singularities lead to new challenges for numerical relativity. New ideas and techniques to analyze and 
compute 
 the asymptotic behavior of multi-physics and multi-scale nonlinear waves have been introduced in the context of classical fluid dynamics but is relevant also for the dynamics of self-gravitating fluids. 
In this short note, we show some numerical results based on structure-preserving numerical algorithms for a few model problems describing multi-scale nonlinear waves on a fixed background. 
We also present a recent proposal 
for defining bouncing cosmologies, which is based on the notion of
 universal scattering maps for gravitational singularities. 
 
This presentation uses material from joint collaborations, especially with B. Le~Floch and G. Veneziano (scattering maps) and with  F. Beyer (Fuchsian algorithm). It is based on a lecture entitled {\sl ``On the scattering laws of bouncing universes''}
given at the {\it Sixteenth Marcel Grossmann Meeting}
 in July 2021.
 It is also based on a lecture  entitled 
 {\sl ``Gravitational singularities, massive fields, and asymptotic localization''}
 given 
 at the Workshop {\it ``Computational Challenges in Multi-Messenger Astrophysics'',} held at 
 IPAM, University of California at Los Angeles, in October 2021. 
 
Only few references are included in this short note and, for our motivations to study 
 bouncing cosmologies and relativistic fluids with complex equation of state, we refer to the review papers 
by Ashtekar \cite{Ashtekar:2009a}, Blaschke \cite{Blaschke}, Font \cite{Font}, as well as the recent papers by
 Ib\'anez et al. \cite{Ibanez} and 
 Wilson-Ewing \cite{Wilson-Ewing}. 
 
An outline is as follows. In Section~\ref{sect-2} our standpoint concerning problems involving viscosity-capillarity waves and other multi-scale interfaces is discussed. 
 In Section~\ref{sect-3}, numerical observations obtained with structure-preserving algorithms are overviewed for fluid problems 
 in the vicinity of gravitational singularities. In Section~\ref{sect-4}, 
 the scattering laws for gravitational singularities are presented and we distinguish between their universal and model-dependent properties. 


\section{Structure of multi-physics and multi-scale waves}
\label{sect-2}

\subsection{Multi-scale wave phenomena}

\paragraph{Systems of balance laws.}

The models of interest are formulated from first principles of continuum physics (that is, conservation laws or more generally balance laws) when suitable physics features are taken into account. 

\bei 

\item Several parameters are often relevant in order to fully describe the fine dynamics of the fluid 
flows of interest: viscosity, surface tension, heat conduction, Hall effect, friction, etc. Various competitive effects take place, due to the presence of several very different scales in, both, the fluid unknowns and the geometry unknowns of the problem. 

\item These nonlinear systems of partial differential equations, in the course of the time evolution, exhibit a formation of 
 interfaces, 
shocks, oscillating patterns, etc. 
Fluids, gases, plasmas, solid materials, etc. all exhibit fine structures that form dynamically.
This is the case of   
 liquid-vapor flows, thin liquid films, 
combustion wave problems, bores in shallow water, astrophysical flows, neutron stars, 
phase transformations (austenite-martensite), etc. 


\item Very often, a {\sl fine-scale structure} (oscillations, turbulence) arises asymptotically (for instance in the vicinity of a singularity hypersurface)
which is challenging to analyze mathematically and to compute numerically. 

\eei 


\

\paragraph{Asymptotic analysis.}

The ideas and techniques developed in the context of fluid dynamics are
 relevant for the modeling of the global dynamics of massive fields (Klein-Gordon), complex fluids, as well as models of 
 modified gravity beyond Einstein gravity. In addition to the shock waves observed in fluid dynamics one must also encompass 
 impulsive gravitational waves and cosmological singularities.

In order to understand the global dynamics of  these 
{\sl scale-sensitive nonlinear waves,} a novel perspective is required which is based on 
an asymptotic analysis of the physical models. 

 \bei 
 
 \item  We restrict attention to the regime where we can extract variables with well-defined limits, even though
  persistent oscillations in some other variables may be observed as the singularity is approached.  

\item  We seek junction laws or scattering laws, which in the case of fluid take the form of jump laws for 
(subsonic) liquid-gas boundaries, for combustion waves, etc. 

\item The interfaces that require additional junction laws are typically 
{\sl under-compressive waves} and turn out to enjoy (saddle) stability properties. 

\eei


\subsection{Regime of small viscosity and capillarity} 

\paragraph{Non-convex equation of state.}

One of the simplest partial differential equation that exhibits small-scale dependent interfaces is the following model in plane-symmetry
formulated as a hyperbolic conservation law with non-convex equation of state $p(\rho) = \rho^3$   
\bel{equa-101}
{ {\del \over \del t} \rho  + {\del \over \del x} \rho^3} 
= \eps \, {\del^2  \over \del x^2} \rho
+ \kappa \, {\del^3  \over \del x^3} \rho, 
\ee   
for a fluid density unknown $\rho  = \rho(t,x)$  (with $t \geq 0$ and $x \in \RR$, say). 
This equation describes certain phase transition phenomena, but also arises as a simplified model of magnetohydrodynamics.  
Here, a  small viscosity coefficient $\eps>0$ and a surface tension/capillarity coefficient $\kappa>0$ are given, the latter representing  
the intermolecular forces between a liquid and its surroundings. 

\begin{figure}
\centerline{\includegraphics[width=10.6cm,height=5.cm]{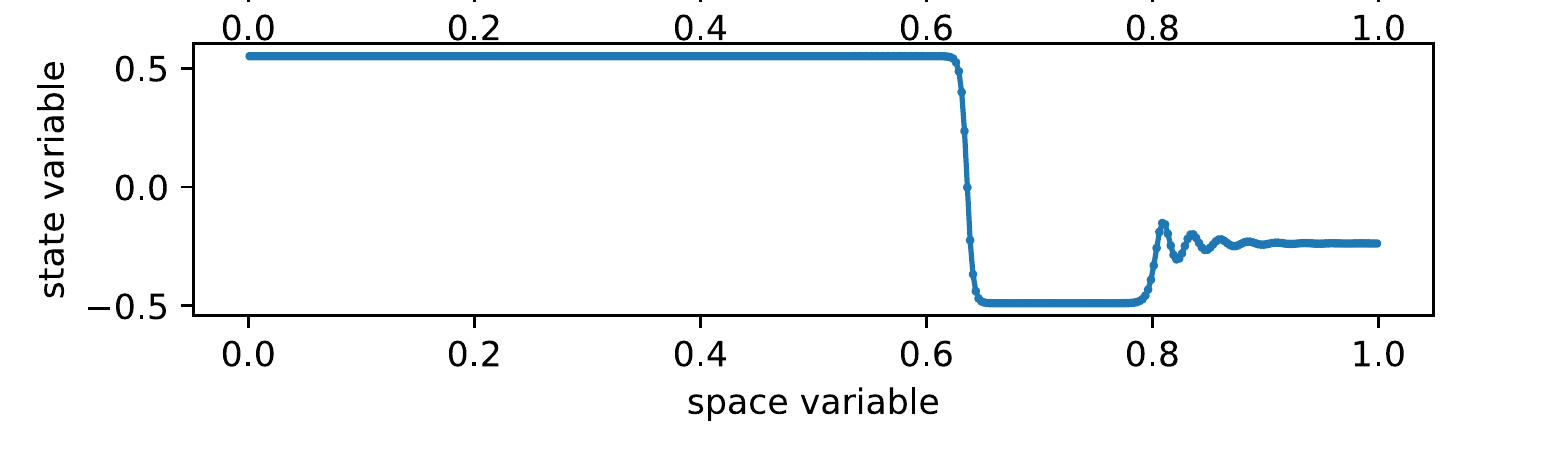}}

\centerline{\includegraphics[width=10.6cm,height=5.cm]{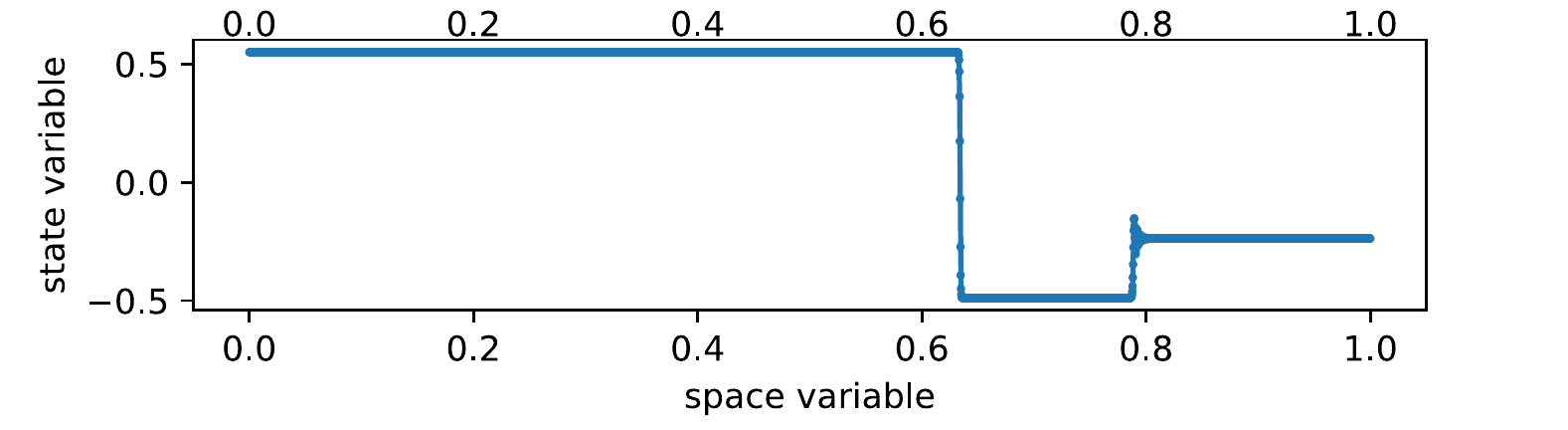}}

\caption{The Riemann problem for a non-convex equation of state.}
\label{Figure---1}
\end{figure}

The Riemann problem consists of an initial value problem associated with a single initial discontinuity (also called the dam breaking problem). Interestingly, in the limit of arbitrarily small $\eps, \kappa \to 0$, the equation \eqref{equa-101}
 exhibits complex wave patterns, which are
 more involved than the ones usually observed with (polytropic perfect, say) compressible fluids.  See the illustration in Figure~\ref{Figure---1}. 
Many scalar wave models exhibit the same features, for instance the so-called thin liquid film model and the so-called
  Benjamin--Bona--Mahony model. 
Let us for instance give here a class of fourth-order models (with $\eps, \kappa, \lambda>0$): 
\be
 {\del \over \del t} \rho  + {\del \over \del x} p(\rho)
= \eps \, {\del^2  \over \del x^2} \rho
+ \kappa \, {\del^3  \over \del x^3}\rho
- \lambda \, {\del^4  \over \del x^4} \rho. 
\ee

\paragraph{Three possible asymptotic regimes.}

In the limit $\eps \to 0$ in \eqref{equa-101}, we should distinguish between three behaviors depending upon the capillarity parameter $\kappa$: 

\bei 

\item $\kappa << \eps^2$ : the viscosity is then dominant, and  no oscillations are observed while a single limit is reached.
The solutions (sometimes referred to as classical weak solutions)  
are characterized by standard Rankine-Hugoniot relations and entropy criteria. 
 
\begin{figure}
 
\centerline{  \includegraphics[width=12.cm, height=6.cm]{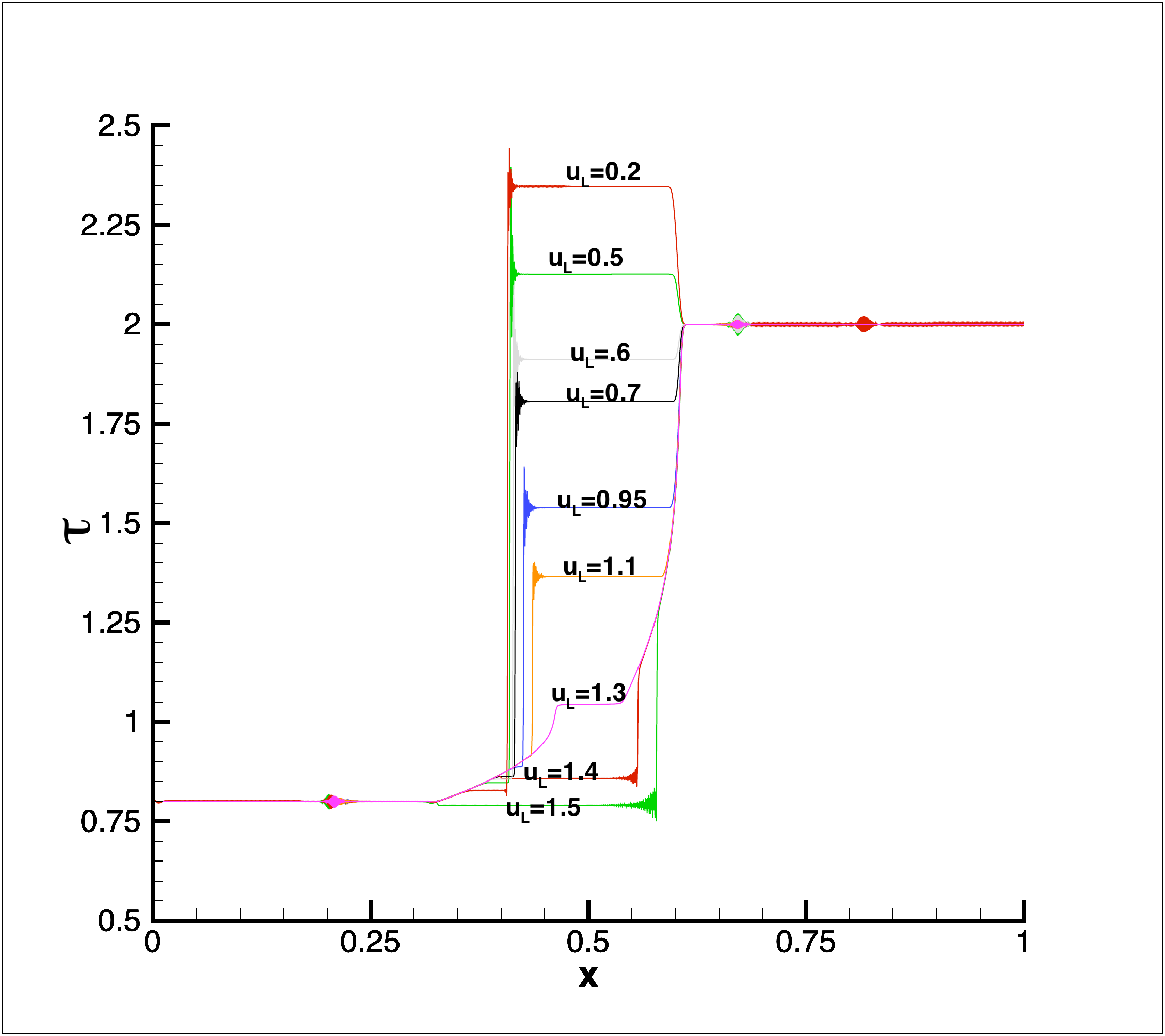}}

\caption{Van der Waals fluids: specific volume for different ratios of the viscosity and capillarity.} 
\label{figure---3} 
\end{figure}

\item $ \kappa = \alpha \, \eps^2$ : this is the {\sl balanced regime} of main interest in the present discussion. The numerical solutions 
converge in  a strong sense, despite some 
mild oscillations in the limit. 
(See again Figure~\ref{Figure---1}.)  
 Interestingly, the limit (now sometimes referred to as a nonclassical weak solution)  
{\sl depends upon the ratio} $\alpha= \kappa/\eps^2$ of the viscosity and capillarity parameters. (See also Figure~\ref{figure---3} for van der Waals fluids.) 

\item $\kappa >> \eps^2$ : in this regime, the effect of the surface tension is dominant
  and wild oscillations arise as $\eps \to 0$.
 There is no well-defined limit, unless weak convergence techniques or a suitable turbulence-like theory are introduced. 
 

\eei 
 
 \
 
  
\paragraph{Wave structure of van der Waals fluids.} 
 
\begin{figure}

\centerline{\includegraphics[width=7.6cm,height=6.cm]{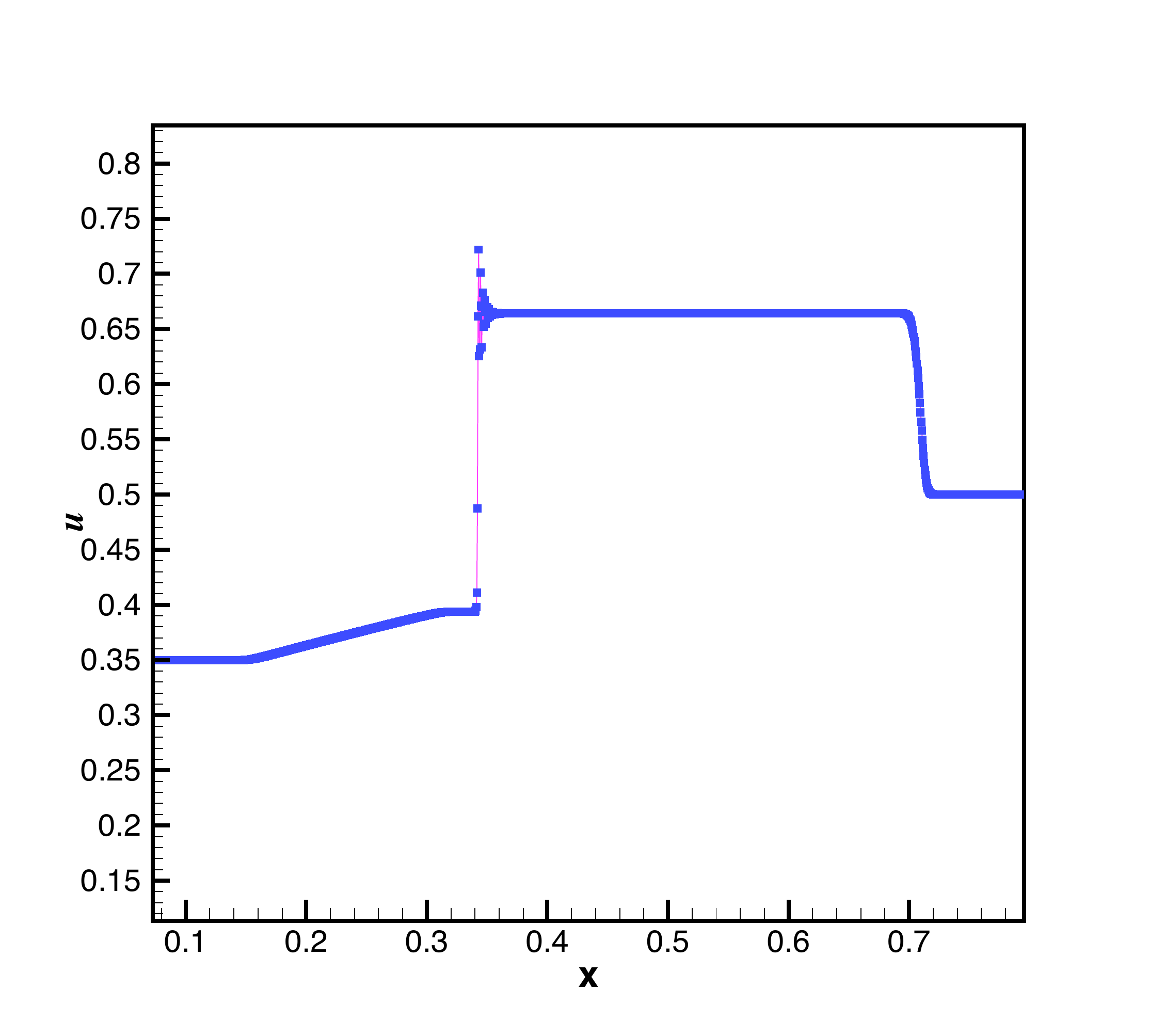}
\includegraphics[width=7.6cm,height=6.cm]{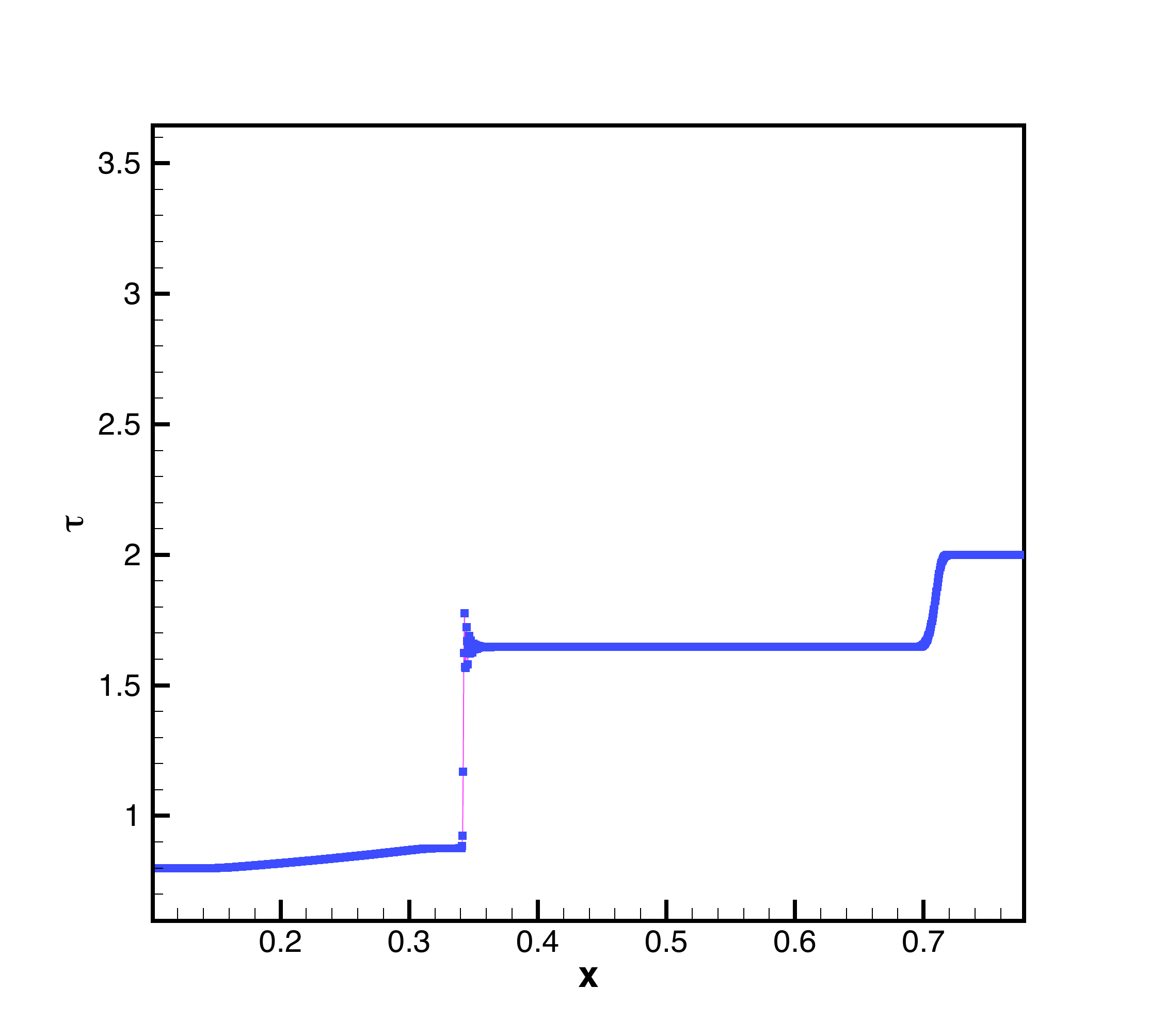} 
}
\caption{Van der Waals fluids: velocity (left-hand) and specific volume (right-hand).} 
\label{figure---2} 
\end{figure}

Let us provide a further illustration with the system of isothermal compressible fluids which is closest to the interest of the GR community. We consider a Van der Waals fluid, in a non-relativistic setup, whose dynamics is modeled by the following
 two conservation laws (continuity equation, conservation of momentum) :  
\bel{equa-105} 
{\del \over \del t} \tau - {\del \over \del x}  u = 0, 
\qquad 
{\del \over \del t}  u + {\del \over \del x}  p(v)
 = \eps \, {\del^2 \over \del x^2}  u  - \kappa \, {\del^3 \over \del x^3} \tau, 
\ee
in which $\tau= 1/\rho$ represents the 
specific volume and $u$ denotes the velocity component for plane-symmetric solutions to the Euler equations. 
The viscosity parameter $\eps$ and the capillarity parameter $\kappa$ are taken to be small. 
We assume that the temperature $T$ remains constant (in a suitable approximation)
and we model liquid-vapor phase transitions by the non-convex pressure law (after normalization) 
\bel{equa-106} 
p(\tau)={ 8T \over {3 \tau-1}} - {3 \over \tau^2}. 
\ee
%
The equations \eqref{equa-105}-\eqref{equa-106}  are nonlinear hyperbolic when $T$ is sufficiently large, but of mixed (hyperbolic-elliptic) type otherwise. 
The typical Riemann wave structure for van der Waal fluids is shown in Figure~\ref{figure---2}. 
 

\paragraph{Proposed standpoint: hidden junction conditions.}

Small-scale dependent, shock waves enjoy a structure that is not properly captured by the standard Rankine-Hugoniot junction conditions of fluid dynamics, even when a suitable entropy condition is enforced. Further analytical work is required which shows
 that the wave structure strongly depends upon 
 varying the ratio surface tension/viscosity $\kappa = \alpha \, \eps^2$ (in the balanced regime). 
 (See again Figure~\ref{figure---3} for van der Waals fluids.) 
 Suitable rules for connecting left- and right-hand state values from both sides of the interface is necessary, beyond the standard Rankine-Hugoniot relations. These {\sl scattering laws} take the form, in this context, of {\sl kinetic functions for interfaces}. 
It is outside the scope of this short Note to explain this notion here, and we refer the reader to the textbook \cite{LeFloch-book}  and 
the lecture notes \cite{PLF-1,PLF-2} and the historical references therein. 
Importantly, based on these scattering maps we need to design structure-preserving algorithms, which 
are typically of front tracking type, or 
of shock capturing type with well-controlled dissipation \cite{PLF-3}. 
The algorithms are designed in order to 
{\sl mimic analytical properties at the discrete level} and this requirement may take very different flavors: 
divergence form, 
 spacelike decay, 
 timelike decay
shock-capturing, 
energy balance laws, 
asymptotics on singularities, etc.

 
\section{Structure-preserving algorithms for fluids near gravitational singularities}
\label{sect-3}

\subsection{Preserving the asymptotic structure on a Kasner background: evolution from the singularity}

\paragraph{Evolution in the vicinity of a gravitational singularity.}

In a joint work with F. Beyer \cite{BL-4},  
I have studied the evolution of a compressible fluid on a Kasner geometry, namely on a 
 spatially homogeneous, anisotropic vacuum background described by the metric 
\be
g = t^{(K^2-)/2} \big( - d t^2+ dx^2 + t^{1-K} dy^2 +t^{1+K} dz^2 \big), 
\qquad
M=(0,+\infty)\times \Tbb^3
\ee
 Here, we set $t \in (0,\infty)$ and $x,y,z \in (0,2\pi)$ and we introduce the parameter $K \in \RR$  and the Kasner exponents
\be
  p_1 = {K^2-1 \over K^2+3}, 
  \quad
  p_2 = {2(1-K) \over K^2+3},
  \quad
  p_3 = {2(1+K) \over K^2+3}. 
\ee
The free parameter $K\in\mathbb R$  is sometimes referred to as the {\sl asymptotic velocity}. Except for the three flat Kasner cases given by
$K=1$, $K=-1$, and (formally) $|K|\to  \infty$, the Kasner metrics $g$ have a curvature singularity in the limit $t\searrow 0$. 
We treat a compressible fluid flow with pressure law 
\be
p= (\gamma-1) \rho \quad \text{ with } \gamma \in (1,2).
\ee


\paragraph{Fuchsian analysis for the Euler equations of compressible fluids.}

In \cite{BL-4} we discovered a {\sl characteristic fluid exponent } defined as 
\be
  \Gamma=\frac 14\left(3 \gamma-2 - K^2 (2 - \gamma)\right) \in (0,1), 
\ee
which compares the geometry and fluid behaviors: 
\bei 

\item $ \Gamma>0$  : this is a {sub-critical regime}  which is  dynamically stable. 

\item $ \Gamma \leq 0$  : this is a {super-critical (or critical) regime} which is    dynamically unstable.  

\eei

\noindent This (positivity/negativity) condition arises by plugging an expansion in power of $t$ for the fluid
 unknown variables of the Euler equations, 
and attempting to validate this expansion in $t$ via the Fuchsian method. 

 
\begin{figure}[t!]

\centerline{
    \includegraphics[height=5.5cm, width=6.cm]{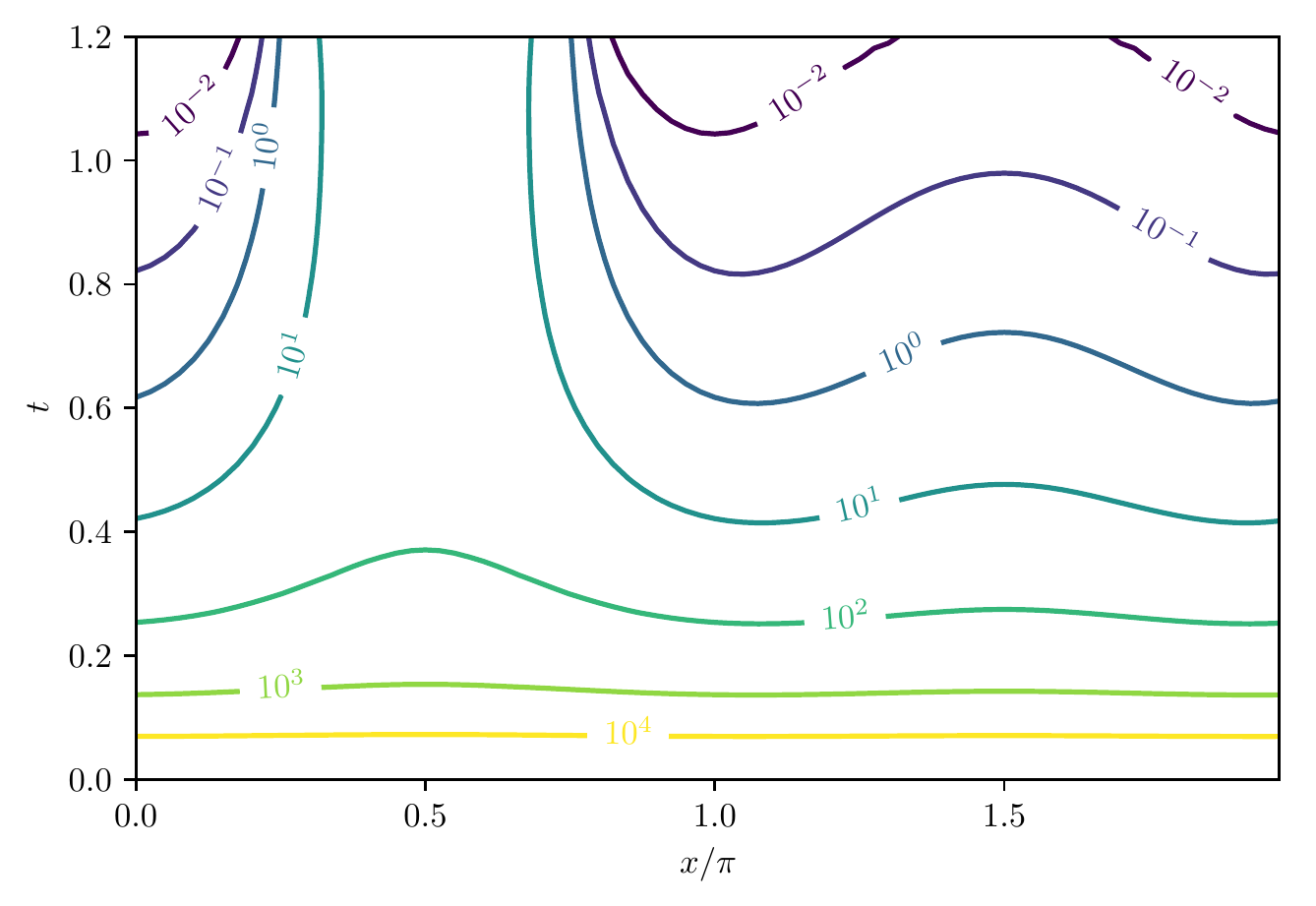}      
%
%
    \includegraphics[height=5.5cm,  width=6.cm]{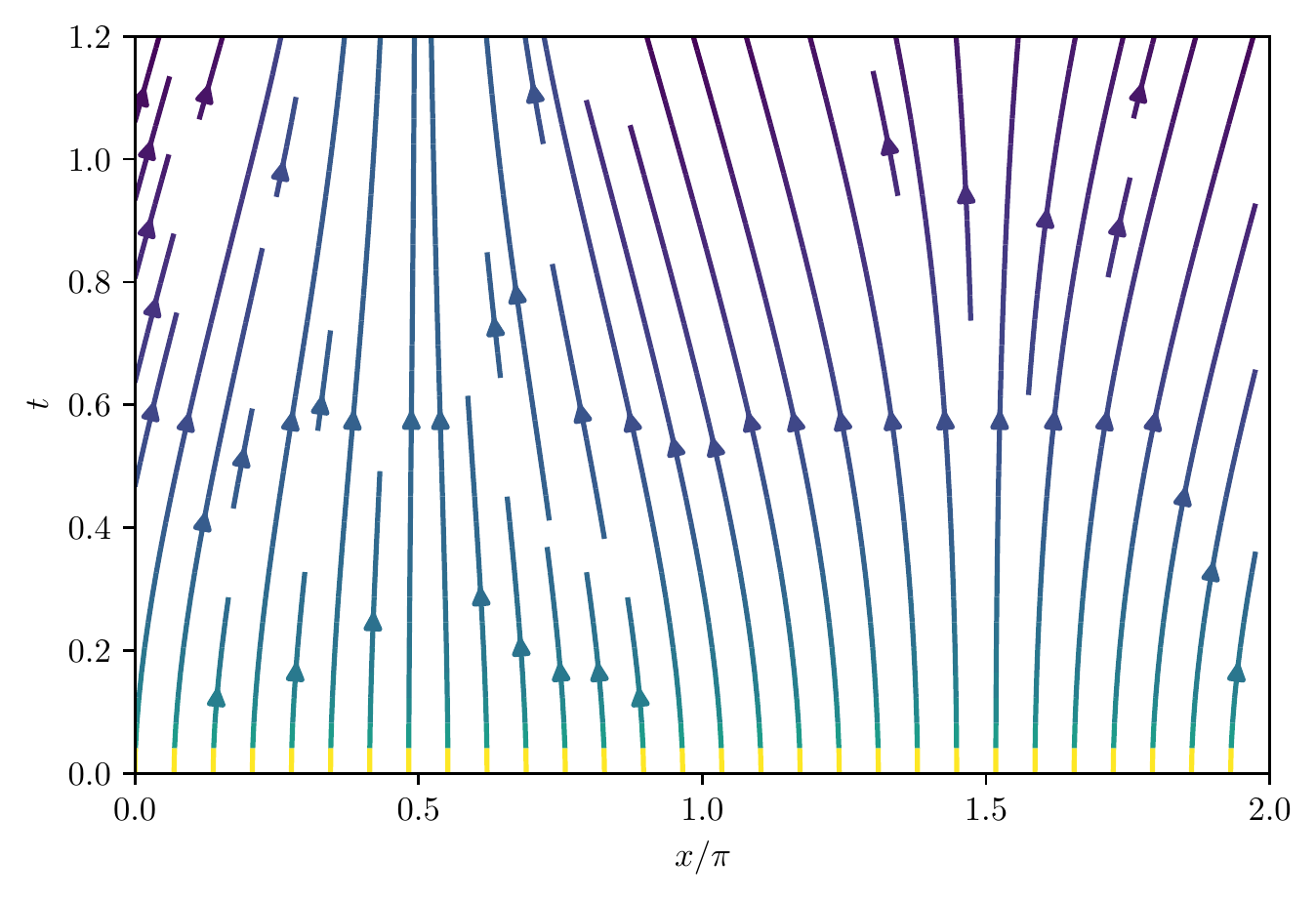}     
}
%
\caption{Evolution of a compressible fluid {\bf from} a gravitational singularity: Kasner background.}
\label{figure-110} 
\end{figure}
 

The Fuschian method concerns the evolution problem {\sl from} the cosmological singularity  $t=0$ (rather than {\sl towards} it). 
We formulate a singular initial value problem of the form 
\be
 B^0 (U, t, x) {\del \over \del t} U + B^1(U,t,x) {\del \over \del x} U = f(U, t, x)
 \ee
with suitable ``singular initial data'' prescribed at $t=0$. Fuchsian-type expansions are then derived from an ODE approximation near the cosmological singularity. In the present work, we focus on the sufficiently regular, shock-free regime for the Euler equations. 


\paragraph{Algorithm preserving the Fuchsian structure.}

We approximate the singular Cauchy problem of Fuchsian type by a sequence of regular Cauchy problems, which we next 
discretize  by the pseudo-spectral method of lines, say $U(t,x) \simeq V(t) = (V_j(t))$ with  
\be
\del_t  V - A  V = h(V,t). 
\ee
A high-order Runge-Kutta discretization is used in time. Importantly, we  introduce {\sl suitably rescaled variables} and next 
 careful study of the numerical error. To this end, we take into account the Fuchsian expansion available near the singularity. We observe that there are 
two sources of approximation error, namely continuum and discrete.

Our proposal in order to get an efficient algorithm    is to {\sl keep the two error sources asymptotically in balance}. 
With this numerical strategy, we can 
demonstrate the {\sl nonlinear stability of the flow} near the cosmological singularity in the sub-critical regime.
See Figure~\ref{figure-110} for a numerical simulation on a Kasner background for a typical evolution from the singularity. 
The fluid density is shown in contour plot while the normalized velocity field is shown in flow lines. 
   
The density $\rho$ unbounded as the time $t \to 0$, and we carefully checked the numerical error 
in order to produce 
quantitative error control. In turn we arrive at 
reliable and accurate algorithm, despite the solutions being {\sl highly singular.} 
We refer the reader to \cite{BL-5} fur further numerical results; the extension to self-gravitating fluids in progress.

%
%

\subsection{Preserving the asymptotic structure on an inhomogeneous FLRW background}

\paragraph{Evolution of fluid with shock waves.}

In a joint work \cite{CGL-1,CGL-2} with Y. Cao and M. Ghazizadeh, I have considered the evolution of a compressible fluid {\sl toward} a cosmological singularity.  
Therein, we consider the Cauchy problem i
isothermal, relativistic compressible flow with linear pressure law $p(\rho) = k^2 \rho$. 
The background geometry is taken to be a contracting (or expanding) inhomogeneous background with torus topology $\mathbb T^2$, corresponding to a FLRW-type cosmological background with small inhomogeneities. For isothermal fluids, 
the Euler equations take the form of a system of two nonlinear hyperbolic balance laws, 
which we treat in one and in two space dimensions under symmetry assumptions. 
Let us focus here on the future-contracting direction ($t<0$ with $t \to 0$) 
and we observe that the mass energy density $\rho \to +\infty$. The expanding direction is also treated in  \cite{CGL-1,CGL-2}.
The challenge is to design a scheme that preserves the structure of the solutions as the singularity is approached and the unknown variables become singular.

\begin{figure}[h!]
\centerline{\epsfig{figure = 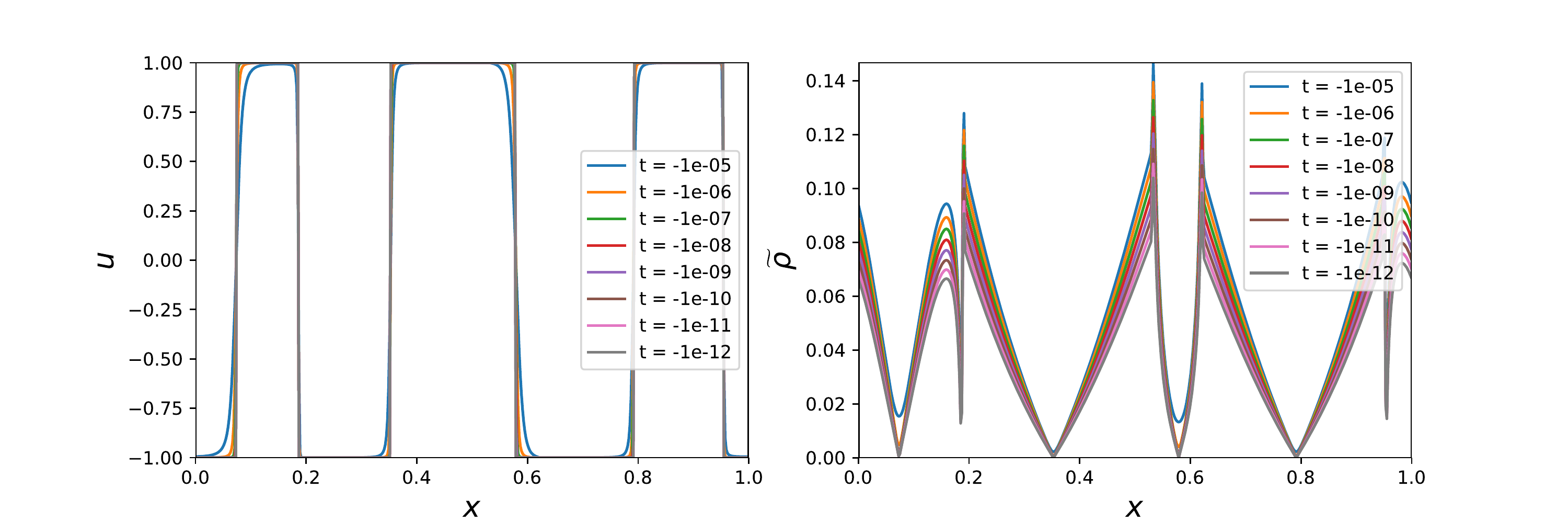,height = 1.8 in, width=15cm} }
\caption{Evolution of a compressible fluid {\bf toward} a gravitational singularity.
Left-hand: velocity scalar. Right-hand: rescaled mass energy density.}
\label {figure-111} 
\end{figure}

We analyze the asymptotic behavior of the fluid variables toward the cosmological singularity, by rewriting first the equations as 
a nonlinear hyperbolic systems on a curved geometry, say  
\be
\del_t U + \del_x F(t,x,U) = H(t,x,U). 
\ee
Two {\sl competitive effects} take place in this problem which concern the contracting geometry 
and the shock propagation. The nonlinear wave interactions give rise 
to a small-scale structure, which in the present setup is simply 
driven by the background geometry. Yet, we can keep in mind 
the analogy with phase transition dynamics which involves multiple scales as explained in Section~2. 
 Extensive numerical experiments were performed 
 in one and in two space dimensions are performed in order to investigate the fine behavior of the solutions in the vicinity of the singularity. 


\paragraph{Structure-preserving methodology.} 

We formulate a shock-capturing scheme in divergence form, based on the  finite volume methodology. The shock-capturing property ensures that the 
shock waves will propagate with the correct speed given by the Rankine-Hugoniot condition. 
We require high accuracy and use a discretization which is 4th-order in time and 2nd-order in space, hence ensure that the numerical solution is oscillation-free.
In order to guarantee the {\sl well-balanced property} we proceed as follows: 

\bei 

\item[$\bullet$] We introduce suitably {\sl rescaled unknowns}  which are directly motivated by a Fuchsian analysis of the problem.

\item[$\bullet$] The discrete form of the balance laws is designed in order to 
 enforce the correct {\sl asymptotic state equations at the discrete level}. 

\item[$\bullet$] In turn , we enforce the {\sl commutation property} 
\be
  \lim_{t \to 0} \lim_{\Delta x \to 0}  U 
=
\lim_{\Delta x \to 0}  \lim_{t \to 0} U
\ee
relating the limits of the numerical solution $U_{\Delta x}(t,x)$

\eei 
 
\noindent 
The typical behavior that we observed {\sl toward} the cosmological singularity are sharp transitions with spikes, and these spikes turn out to be well capured by our numerical method in the asymptotic limit $t \to 0$. 
See the plots of the 
rescaled velocity component $u$ and rescaled density $\widetilde \rho$. See Figure~\ref{figure-111} for 
the evolution of a compressible fluid {\bf toward} a gravitational singularity.


\section{On the scattering laws of bouncing universes}
\label{sect-4}

\subsection{Seeking a formulation of bouncing laws}

\paragraph{Bouncing cosmologies.}

In a joint work in collaboration with B. {Le~Floch}  (Paris) 
 and G. Veneziano  { (Geneva)} I considered gravitational singularities in the context of 
 the  Einstein equations coupled to a self-gravitating scalar field. 
 More generally, our ideas apply to a stiff fluid or a compressible fluid. 
We are interested in analyzing bouncing cosmologies involving a contracting phase and an expanding phase of evolution, connected across a singularity hypersurface. Our analysis takes place in the vicinity of such a singularity hypersurface, and we focus on the jump relations that should hold at that interface. 
Recall that in presence of strong fields or away from the near-Minkowski regime of evolution, 
Penrose and Hawking's singularity theorems (or incompleteness theorems) suggest that gravitational 
singularities (or other incompleteness phenomena) 
are expected to occur, 
although very little is known on the actual asymptotic behavior beyond the classical BKL conjecture. 


\paragraph{Physics literature on bouncing.}

We build upon a large literature on bouncing cosmologies by Penrose, Tod, L\"ubbe, Turok, Barrow, and many others. We emphasize that existing proposals are based on the analysis of symmetric spacetimes and/or the formulation of
 special junctions. In order to define the bouncing, 
 a large number of proposals are found in the literature, including the pre-Big Bang scenario (Gasperini, Veneziano, etc.), 
a variety of models of modified gravity-matter (Brandenberger,  Chamseddine, Cotsakis, 
  Mukhanov, Peter, Steinhardt, Turok, etc.), and 
the theory of loop quantum cosmology (Asthekar, de Cesare, Gupt, Pawlowski, 
  Singh, Wilson-Ewing, etc.). 

\


\paragraph{Proposed standpoint.}

Our aim is to propose a formulation of the problem of junction at singularities and to perform a systematic investigation of such junction conditions. This is motivated by the ideas discussed in Section~2. 
We expect to derive classes of physically meaningful junction conditions 
which depend upon certain (limited) degrees of freedom and constraints at the singularity. Interestingly, based on the scattering maps we propose we can also define and construct classes of 
cyclic spacetimes. Here, we only outline a selection of our results and  refer the reader to \cite{LLV-1,LLV-2,LLV-3} for the complete definitions and further material.

We study the class of (past, future) {\sl singularity data} denoted by $(g^\pm, K^\pm, \phi_0^\pm, \phi_1^\pm)$ and
 defined after a suitable rescaling (see below). We then introduce a
{\sl singularity scattering map} 
$
(\gmoi, \Kmoi, \phimoi_0, \phimoi_1) \mapsto (g^+, K^+, \phi_0^+, \phi_1^+)
$
in order to cross the singularity hypersurface. Our main contribution is a full classification of the possible bouncing conditions. 
We are able to distinguish between {\sl universal} scattering laws and {\sl model-dependent} scattering maps.

We thus provide a flexible framework and a classification which 
 uncovers all possible classes of junctions that are geometrically and physically meaningful. We also 
 exhibit several sub-classes of particular interest, namely the conformal/non-conformal maps, the spacelike/null/timelike maps, and we 
 encompass different models of matter
 (scalar field, stiff fluid, compressible fluid). 
In the course of establishing our classification, we also discover three universal laws  that put 
constrains on the macroscopic aspects of spacetime bounces, 
{\sl regardless of their origin from different microscopic corrections}. 
Our classification provides one with a guide in order to 
uncover relevant structures, described by scattering maps associated with specific theories. 


\subsection{Proposed formulation of the bouncing problem} 

\paragraph{The field equations.}

We consider a local ADM formulation near a singularity hypersurface, based on a Gaussian foliation (local patch) 
$\Mcal^{(4)} = \bigcup_{\tau \in [\tau_{-1}, \tau_1]} \Hcal_\tau$ 
by spacelike hypersurfaces  (diffeomorphic to $\Hcal_0$). We write 
\be 
g^{(4)} = \big(g^{(4)}_{\alpha\beta}\big)
= - d\tau^2 + g(\tau) \quad \qquad
 g(\tau) = g_{ij}(\tau) dx^i dx^j, 
\ee
and we impose Einstein's evolution equations for the unknown metric $g$ and the unknown extrinsic curvature $K$ 
\be
\del_\tau g_{ij}  = - 2 \, K_{ij}
\qquad \qquad 
\del_\tau K^i_j = \Tr(K)  K^i_j + R^i_j - 8 \pi \, M^i_j. 
\ee
Here, $M^i_j =  {1 \over 2} \rho g^i_j + T^i_j - {1 \over 2} \Tr(T)  g^i_j$ is the matter contribution. 
These equations are supplemented with Einstein's constraint equations
\be
R + |K|^2 - \Tr (K^2) = 16 \pi \rho
\qquad 
\nabla_i K^i_j  - \nabla_j (\Tr K) = 8 \pi J_j, 
\ee
which are nonlinear elliptic equations. Finally, we assume that the energy-momentum tensor is given by 
a massless scalar field
$\phi$ whose evolution is given by  the wave equation 
\be
\Box_{g^{(4)}} \phi = 0. 
\ee 

Near a singularity hypersurface, it is standard to rely on the Fuchsian method (Baouendi, Goulaouic, Rendall, Isenberg, Moncrief, etc.): 

\bei

\item[--] We solve locally from $\tau=0$ toward the past ($\tau<0$) or the future ($\tau>0$). 

\item[--] We apply the  so-called ``velocity dominated'' Ansatz, which tells us that
 in the gauge under consideration 
 all spatial derivatives (except for
 the momentum equations) can be neglected. 

\item[--] In turn, we must solve a system of nonlinear coupled differential equations in the time variable with 
nonlinear 
source-terms (containing spatial derivatives) 
and by a suitable iteration scheme we generate solutions to the full Einstein equations. 

\eei 

 
\paragraph{Singularity data and asymptotic profiles.}

A $3$-manifold $\Hcal$ is fixed throughout and represents the singularity hypersurface. 

\bei 

\item[1.] An {\bf asymptotic profile} associated with data $(\gmoi, \Kmoi, \phimoi_0, \phimoi_1) \in \Ibf(\Hcal)$ is the flow on $\Hcal$ 
\be
\aligned
\tau \in (-\infty,0) & \mapsto \big(g^*, K^*, \phi^*\big) (\tau), 
\hskip1.cm
&& g^*(\tau) = |\tau|^{2 \Kmoi} \gmoi, 
\\
K^*(\tau) & = {-1 \over \tau} \Kmoi, 
&&
\phi^*(\tau) = \phimoi_0 \log|\tau|  + \phimoi_1, 
\endaligned
\ee
in which $ |\tau|^{2 \Kmoi} $ is defined by exponentiation. 

\item[2.] A {\bf singularity initial data set} $(\gmoi, \Kmoi, \phimoi_0, \phimoi_1)$, 
consists of two tensor fields (rescaled metric and extrinsic curvature)
and two scalar fields defined on $\Hcal$. 

\item[3.] The {\bf asymptotic version of the Einstein constraints} reads as follows: 
\be
\aligned
& 
\quad \text{Riemannian metric }            && \gmoi =(g^-_{ij})  \, \text{ on } \Hcal.
\\
&
\quad \text{CMC symmetric $(1,1)$-tensor } && \Kmoi = (K_i^{-j}) 
\text{ with } \Tr(\Kmoi) = 1  \text{ on } \Hcal.
\\
& 
\quad \text{Hamiltonian constraint}  && 1 - | K^-|^2
= 8\pi \, (\phimoi_0)^2   \text{ on } \Hcal. 
\\
& 
\quad \text{momentum constraints}  &&
\Div_{g^-} (K^-)
= 8 \pi \, \phimoi_0 d\phimoi_1  \text{ on } \Hcal.
\endaligned
\ee

\eei  
 
\noindent We denote by $\Ibf(\Hcal)$ the set of space of all singularity data. 


\paragraph{Junction conditions.} 

The notion of a {\sl scattering map} arises naturally from earlier considerations about fluid interfaces. 
We think of a singularity hypersurface as a fluid-like interface between two ``phases'', across which the geometry and the matter field encounter a ``jump''. Small-scale physics phenomena are not directly modeled at this stage. 
We are only interested in the ``average'' effect rather than the detailed physics that may take place ``within'' this interface. 
This is a standard strategy in fluid dynamics and material science in presence of phase transition phenomena, especially 
when some (micro-scale) parameters (viscosity, surface tension, heat conduction, etc.,)
can be neglected in the modeling. Macro-scale effects are captured by jump conditions  such as 
Rankine-Hugoniot, kinetic relations, Dal~Maso-LeFloch-Murat's paths, etc. 
For instance, kinetic relations in material science (martensite-austenite) 
and in two-phase liquid-vapor flows have been extensively studied. In this context, hypersurfaces are timelike;
 in the present discussion of gravitational singularities we focus our attention on spacelike hypersurfaces and refer the reader to \cite{LLV-2} for the treatment of timelike singularity hypersurfaces.


\paragraph{Bounce based on a singularity scattering map.} 

A (past-to-future) {\bf singularity scattering map} on $\Hcal$ is defined as a diffeomorphism-covariant map on $\Ibf(\Hcal)$
\be
\Sbf: \Ibf(\Hcal)  \ni \big(\gmoi, \Kmoi, \phimoi_0, \phimoi_1\big)  
\mapsto 
\big(\gpoi, \Kpoi, \phipoi_0, \phipoi_1\big) \in \Ibf(\Hcal)
\ee
satisfying the {\sl ultra-locality property}: for all $x \in \Hcal$  
\be
\Sbf(\gmoi, \Kmoi, \phimoi_0, \phimoi_1)(x) \text{ depends only on } (\gmoi, \Kmoi, \phimoi_0, \phimoi_1)(x)
\ee
Thanks to the locality condition, it is natural to identify singularity scattering maps~$\Sbf$ on all $3$-manifolds and  suppress the dependence on~$\Hcal$.


The map $\Sbf$ is said to be a {\bf tame-preserving map}
if it preserves positivity, in the sense that
\be
\text{if } \Kmoi > 0 \text{ then } \Kpoi > 0 , \text{ where } \Kpoi \text{ is defined from the image of $\Sbf$.}
\ee
It is said to be a {\bf rigidly-conformal map} if 
\be
\text{$\gpoi$ and $\gmoi$ only differ by a conformal factor.}
\ee


\paragraph{Observations.}

We emphasize that the asymptotic profiles with $\Kmoi,\Kpoi>0$ describe a ``bounce'' at which 

\bei 

\item[--]  volume element decreases to zero as $\tau\to 0^-$, and 

\item[--] then increases back to finite values for $\tau>0$. 

\eei 

\noindent For further notions and constructions of {\sl cyclic spacetimes} containing many singularity hypersurfaces, we refer to \cite{LLV-2} and, in this short Note, we focus on the issue of defining a proper junction at the bouncing.

The regime of quiescent singularities $K>0$ is motivated by the absence of BKL oscillations in this case (named after Belinsky, Khalatnikov, and Lifshitz). This quiescent regime exhibits a monotone behavior which has received a lot of attention by mathematicians in recent years (Rendall, Andersson, Lott, Fournodavlos, Luk, Rodnianksi, Speck, etc.). 


\subsection{Classification of bouncing laws}
 
\paragraph{Rigidly conformal maps.}

We discover that {\sl only two classes} of ultra-local spacelike and rigidly conformal, singularity scattering maps exist 
for self-gravitating scalar fields, namely:  
 
\bei 

\item The maps describing {\bf isotropic rigidly conformal bounces} $\Sbf^\text{iso, conf}_{\lambda,\varphi}$: 
\be
g^+ = \lambda^2 \gmoi,
\qquad
K^+ = \delta/3, 
\qquad
\phi_0^+ =  1/\sqrt{12\pi}, 
\qquad
\phi_1^+ =  \varphi,  
\ee
 parametrized by a conformal factor $\lambda=\lambda(\phimoi_0,\phimoi_1,\det\Kmoi)>0$  and a constant~$\varphi$. 


\item The maps describing {\bf non-isotropic rigidly conformal bounces}  $\Sbf^\text{ani, conf}_{f,c}$: 
\be
\aligned
&  g^+ = c^2 \mu^2 \gmoi, 
\qquad
&&
K^+ = \mu^{-3}(\Kmoi - \delta/3) + \delta/3, 
\\
& 
\phi_0^+ =  \mu^{-3} \phimoi_0 / F'(\phimoi_1), 
\qquad
&&
\phi_1^+ = F(\phimoi_1),
\endaligned
\ee
parametrized by a constant ${c>0}$ and a function $f\colon\RR\to [0, +\infty)$: 
\be
\mu(\phi_0, \phi_1) 
=  \big(1+12\pi (\phi_0)^2 f(\phi_1) \big)^{1/6}, 
\, 
F(\phi_1)= \int_0^{\phi_1} (1+f(\varphi))^{-1/2} d\varphi.
\ee

\eei
%

 
\paragraph{General classification.} 
 
{\sl Only two classes} of ultra-local spacelike singularity scattering maps exist for self-gravitating scalar fields, which we denote by 
$\Sbf^\text{iso}_{\lambda,\varphi}$ (isotropic bounce) 
and
 $\Sbf^\text{ani}_{\Phi,c}$ (non-isotropic bounce). 
Now, $\lambda$ is a two-tensor, $\Phi$ a ``canonical transformation'', $c$ a constant. For the detailed expressions we refer the reader to \cite{LLV-2}. 


\paragraph{Three universal laws of quiescent bouncing cosmology.}

Interestingly our conclusions can be restated in terms of universal laws obeyed by any ultra-local bounce, as follows. 
\bei 

\item {\bf First law. Scaling of Kasner exponents}: 
\be
\aligned
& \text{ There exists a (dissipation) constant $\gamma\in\RR$ such that}
\\
&
 |g^+ |^{1/2} \Kcirc^+ = - \gamma \, |g^- |^{1/2} \Kcirc^-
\endaligned
\ee
for the spatial metric~$g$ in synchronous gauge with volume factor $|g|^{1/2}$, where 
$\Kcirc$~denotes the traceless part of the extrinsic curvature (as a $(1,1)$ tensor). 


\item {\bf Second law. Canonical transformation of matter:}    
by denoting the conjugate matter momentum by $\pi_\phi \sim \phi_0$, 
\be
\aligned
& \text{there exists a nonlinear map } \Phi\colon(\pi_\phi,\phi)^- \mapsto (\pi_\phi,\phi)^+
\\
& \text{{preserving the volume form }in the phase space } d\pi_\phi\wedge d\phi 
\\
&   \text{{and depending solely on the scalar invariant} $\textbf{det}(\Kcirc_-)$.}
\endaligned
\ee


\item {\bf Third law. Directional metric scaling:}
\be
\aligned
& 
g^+= \exp\bigl(\sigma_0 + \sigma_1 K + \sigma_2 K^2\bigr) g^-, 
\\
& \text{which is a nonlinear scaling in each proper direction of~$K$.}
\endaligned
\ee
We have either $\gamma=0$ in the isotropic scattering and no restriction $\sigma_0,\sigma_1,\sigma_2$, or else 
  $\gamma\neq 0$ for non-isotropic scattering and explicit formulas are available in terms of $\Phi,\gamma$.

\eei


\subsection{Ongoing work} 

Our next task will be the derivation of scattering maps associated with specific theories taking 
small-scale physical modeling into account. Mathematical investigations as well as numerical investigations are required. 
We emphasize that our methodology is based on a class of {\sl general}  spacetimes without symmetry restrictions. 
Earlier approaches relied on classes of symmetric spacetimes or on special junction proposals, and therefore
could not single out 
the three universal laws we have derived. The recent developments overviewed in the present Note
 suggest to perform numerical simulations of cyclic spacetimes with a bounce using some of the techniques discussed in Sections~2 and 3. We refer the reader to \cite{LLV-1,LLV-2,LLV-3} and to \cite{LL-2}.  


\end{document}